# Abrupt declines in tropospheric nitrogen dioxide over China after the outbreak of COVID-19

Fei Liu[1,2]*, Aaron Page[3], Sarah A. Strode[1,2], Yasuko Yoshida[2,4], Sungyeon Choi[2,4], Bo Zheng[5], Lok N. Lamsal[1,2], Can Li[2,6], Nickolay A. Krotkov[2], Henk Eskes[7], Ronald van der A[7], Pepijn Veefkind[7,8], Pieternel Levelt[7,8], Joanna Joiner[2†], Oliver P. Hauser[9†]

[1]Universities Space Research Association (USRA), Columbia, MD 21046, USA.

[2]NASA Goddard Space Flight Center Laboratory for Atmospheric Chemistry and Dynamics, Greenbelt, MD 20771, USA.
[3]Department of Management, University of Exeter, Exeter EX4 4PU, UK.
[4]Science Systems and Applications and Applications, Inc., Lanham, MD 20706, USA.
[5]Laboratoire des Sciences du Climat et de l'Environnement, CEA-CNRS-UVSQ, Gif-sur-Yvette, UMR 8212, France.
[6]Earth System Science Interdisciplinary Center, University of Maryland, College Park, MD 20740, USA.
[7]Royal Netherlands Meteorological Institute (KNMI), De Bilt 3731 GA, the Netherlands.
[8]Delft University of Technology, Delft 2628 CD, the Netherlands.
[9]Department of Economics, University of Exeter, Exeter EX4 4PU, UK.

*Correspondence to: Fei Liu (fliu@usra.edu); †Contributed equally.

**Abstract:** China's policy interventions to reduce the spread of the coronavirus disease 2019 have environmental and economic impacts. Tropospheric nitrogen dioxide indicates economic activities, as nitrogen dioxide is primarily emitted from fossil fuel consumption. Satellite measurements show a 48% drop in tropospheric nitrogen dioxide vertical column densities from the 20 days averaged before the 2020 Lunar New Year to the 20 days averaged after. This is 20% larger than that from recent years. We relate to this reduction to two of the government's actions: the announcement of the first report in each province and the date of a province's lockdown. Both actions are associated with nearly the same magnitude of reductions. Our analysis offers insights into the unintended environmental and economic consequences through reduced economic activities.

**One Sentence Summary:** Chinese COVID-19 policies relate to fuel use.

In December 2019, a respiratory disease, coronavirus disease 2019 (COVID-19), emerged in Wuhan City, Hubei Province, China (*1*). COVID-19 has since spread worldwide causing tens of thousands of deaths (*2*). To combat the spread of COVID-19, the Chinese government sealed off several cities reporting large numbers of infected people, including Wuhan, starting January 23, 2020; this included halting public transportation and closing local businesses. These prevention efforts quickly expanded nationwide. The policy announcements and restrictions, applied at an unprecedented scale, have implications for the Chinese environment and the economy that we quantitatively evaluate in this paper. In particular, we use satellite nitrogen dioxide ($NO_2$) measurements to monitor changes in fossil fuel usage, related to economic activity, over China following the outbreak of COVID-2019. Nitrogen oxides ($NO + NO_2 = NO_x$), emitted during high temperature combustion, are relatively short-lived in the atmosphere (lifetimes of the order of hours near the surface), and therefore remain relatively close to their sources (*3*). $NO_2$ tropospheric vertical column density (TVCD) retrieved from backscattered solar radiation, such as from the Ozone Monitoring Instrument (OMI; *4*), has been widely used to monitor both long term and short-term changes in fuel consumption (*5, 6*). OMI's successor, the Tropospheric Monitoring Instrument (TROPOMI; *7*) offers a higher spatial resolution measurement of $NO_2$ TVCD.

We observe substantial reductions of $NO_2$ TVCD after the 2020 Lunar New Year (LNY) on January 25, 2020. Figure 1 shows 20-day averages of OMI $NO_2$ TVCD before, during and after the 2020 LNY (hereafter referred to as the "pre", "peri" and "post" periods). An average reduction of 48% in $NO_2$ TVCD over China is observed from pre to peri periods. Consistency in the trends of retrieved $NO_2$ TVCD is found between OMI and its successor TROPOMI (Figure S1). A reduction in $NO_2$ TVCD is typically observed during LNY because most Chinese factories shut down for the holiday and the traffic volumes decrease, resulting in a decrease in fuel consumption and thus $NO_x$ emissions. OMI $NO_2$ TVCD shows an average pre to peri decline of 26% from data covering the 2015 to 2019 period (Fig. S2). Similarly, TROPOMI shows a reduction of 33% in 2019 (Fig. S3). This suggests that the observed reduction in 2020 far exceeds the typical holiday-related pre to peri period reduction.

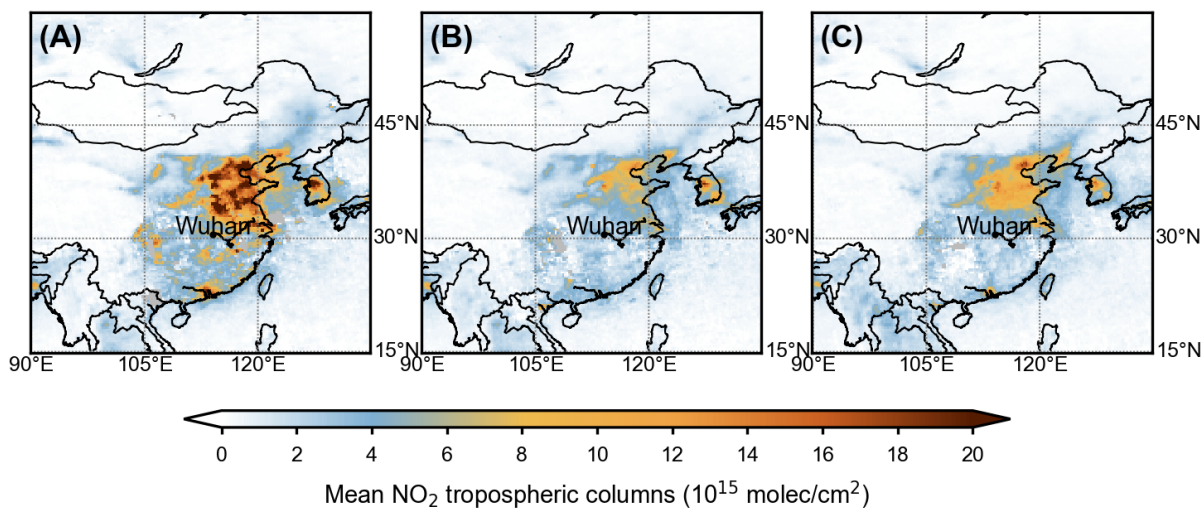

**Fig. 1. Average OMI tropospheric $NO_2$ vertical column densities over China in 2020.** (A) -20 to -1, (B) 0-19, and (C) 20-39 days relative to the 2020 Lunar New Year.

Consistent with the 2015–2019 data, the 2020 $NO_2$ TVCD 7-day moving averages show a significant reduction during approximately the two weeks leading up to LNY and reach a minimum around LNY, consistent with the gradual shutdown of factories before the holiday (Fig. 2). In prior years, a rebound of $NO_2$ TVCD usually begins around 7 days after LNY, marking the end of the holiday season. OMI and TROPOMI (Fig. S4) $NO_2$ TVCDs show similar temporal patterns prior to 2020 with a clear reduction before LNY and an increase shortly thereafter. However, while the 2020 data show similar initial declines in the week leading up to LNY, we do not observe the typical uptick in $NO_2$ TVCDs starting the week after the LNY as in previous years (Fig. 2). OMI (and TROPOMI) $NO_2$ TVCDs show a longer period of low values near the minimum. Note that the 2020 data are generally lower than previous years, probably reflecting in part the effects of China's clean air policies that require installation of denitrification devices for all coal-fired power plants and cement plants (*8*).

To rule out the possibility that the large $NO_2$ TVCD decreases observed in 2020 may be driven by changes in the meteorological conditions affecting local $NO_x$ chemistry and $NO_x$ transport, we use Goddard Earth Observing System version 5 Chemistry-Climate Model (GEOS-CCM; *9*) simulations with constant emissions. We find the simulated effects of meteorology on $NO_2$ TVCD small as compared with the prolonged $NO_2$ reduction we observe from the pre to peri period (Fig. S5). The simulation with constant emissions shows many areas with increases from the pre to peri periods (Fig. S6). This suggests that in many areas the actual decrease in $NO_x$ emissions may be larger than what is inferred from the observed $NO_2$ TVCDs.

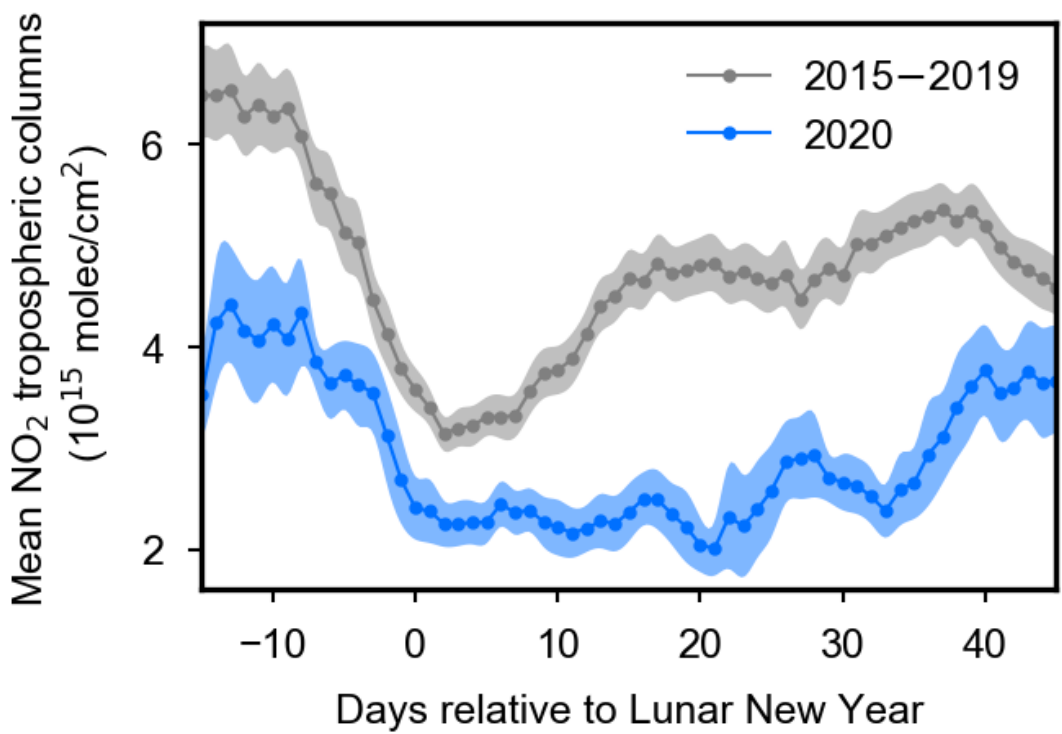

**Fig. 2. Daily variations in 7-day moving averages of OMI NO$_2$ TVCDs over China.** Shading shows standard error of the mean. Points are plotted at the midpoint of the 7-day moving average. Note that we account for the annually varying dates of the Lunar New Year.

Breaking these results down by sectors provides insights into the sources of reduction. All sectors experienced dramatic NO$_2$ reductions. We compute 7-day running averages for all OMI observations within 0.25° gridboxes that contain large power plants or other industrial plants with reported NO$_x$ emissions > 5 Gg/yr (Fig. S7). OMI NO$_2$ TVCD averages for gridboxes containing power plants and those for other industrial plants show similar temporal variations as the national average (Fig. S8). This suggests that measures to reduce COVID-19 spread affected power generation as well as industrial production including steel, iron, and oil. Direct NO$_2$ reductions from transportation are indicated by the visually reduced TROPOMI $NO_2$ TVCDs along the China National Highways (Fig. S9).

We next explore how COVID-19 policy interventions (most of which happened to coincide with the 2020 LNY) are associated with reductions in $NO_2$ TVCD. First, we consider the announcements the government made to the public (Table S1). Once the government publicly reports that a COVID-19 case has been confirmed in a province, the public in that province might choose to reduce their exposure to others (e.g., stay at home, work from home and/or travel less). In that case, we would expect a reduction in $NO_2$ TVCD following the announcement of the first case in each province. This is indeed what we find, after taking previous years' $NO_2$ TVCD and variation across provinces into account (see Eq. S1): following the report of the first case in each province, OMI $NO_2$ TVCD declined by about 27% (coeff = -1.383, $p$ = 0.002, Table 1 Col. 1).

The second policy intervention is more invasive: the government took decisive action to further reduce the spread of the virus by limiting the mobility of citizens and locking down entire provinces; on average, lockdowns occurred 3.7 days after the report of the first case. We would expect that a lockdown would be followed by a reduction in travel as well as business activity, which in turn should lead to reductions of $NO_2$ TVCD. Our model (Eq. S2) shows that OMI $NO_2$ TVCD reduces by 24% following the lockdowns (coeff = -1.134, $p$ < 0.001, Table 1 Col. 2).

Finally, we consider the two policies jointly (Eq. S3). We find that both the announcement of the first case reported as well as the lockdown are associated with a reduction in $NO_2$ TVCDs in each province (Table 1 Col. 3). These results suggest that the effect of the announcement is about as large (16%; coeff = -0.851, *p* = 0.043) as the effect of the lockdown (15%; coeff = -0.752, *p* < 0.001). All results are qualitatively similar using TROPOMI (Table S2).

**Table 1. Effects of the government policies on $NO_2$ tropospheric vertical column density (TVCD)**

|  | Outcome variable: | | |
|---|---|---|---|
|  | $NO_2$ TVCD ($10^{15}$ molec/cm$^2$) | | |
|  | (1) | (2) | (3) |
| First case announced in province, $\beta$ | -1.383** |  | -0.851* |
|  | (0.409) |  | (0.401) |
| Lockdown of province, $\lambda$ |  | -1.134*** | -0.752*** |
|  |  | (0.226) | (0.158) |
| Average $NO_2$ TVCD 2015-2019, $\delta$ | 0.0001 | 0.004 | -0.002 |
|  | (0.019) | (0.018) | (0.019) |
| Constant, $\alpha$ | 5.122 | 4.660 | 5.176 |
| Number of observations | 968 | 968 | 968 |
| $R^2$ | 0.547 | 0.548 | 0.554 |
| Adjusted $R^2$ | 0.533 | 0.534 | 0.539 |

*Note.* $NO_2$ TVCD is based on OMI. We use a fixed-effects model (Eqs. S1-3) with first case announced and lockdown coded as binary indicator variables. We control for the average 2015–2019 OMI $NO_2$ TVCDs to adjust for seasonal variation and include provinces' fixed-effects to adjust for geographical variation. The "Constant" term is the average province fixed-effect used as a baseline to compare the relative effect of the policy interventions. All standard errors (shown in parentheses) are clustered at the province level. * $p < 0.05$, ** $p < 0.01$, *** $p < 0.001$.

$NO_2$ reductions are closely related to improvements in air quality (*10*). Under normal circumstances, many Chinese cities have poor air quality that reduces life quality and expectancy (*11*). During the COVID-19 crisis, $NO_2$ pollution was additionally reduced by ~20% for a period of between 30 and 50 days. While temporary, these substantial reductions in air pollution may have positive health impact for lives in otherwise heavily polluted areas (*12*). This unusual period offers a rare counterfactual of a potential society which uses substantially less fossil fuels and has lower mobility (*13*).

While this research provides an early insight into the $NO_2$ changes in China in early 2020, our findings are not without limitations. Because the relationship between $NO_2$ TVCD and $NO_x$ emissions is not strictly linear, the analysis of $NO_2$ TVCD provides a qualitative description of changes in $NO_x$ emissions. Accurately quantifying the changes in $NO_x$ emissions (*14*) is beyond the scope of this initial assessment.

Our results suggest that the announcement of the first case was followed by a reduction in $NO_2$ emissions, with a further reduction following the actual lockdown. However, it is important to note that these results do not suggest that the mobility restrictions did not have a critical impact. Indeed, recently published work suggests that the travel restrictions in China reduced the spread of the disease by up to 80% by mid-February, in particular internationally (*15*). In line with our results is the finding that human mobility was reduced early on during the outbreak (*16*) and may in part have started as early as the first case announcements, with additional reductions through lockdowns.

**Acknowledgments:** The authors thank the algorithm, processing, and distribution teams for the OMI and TROPOMI data sets used here. The authors thank Dr. Luke Oman for helping to set up the GEOS-CCM model runs and emissions. **Funding:** Funding for this work was provided in part by NASA through the Aura project data analysis program and the ACMAP managed by Ken Jucks, Barry Lefer, and Richard Eckman, who the authors acknowledge for their continued support; **Author contributions:** Conceptualization and Methodology: F. L., A. P., J. J., O. P. H.; Formal Analysis: F. L., A. P., O. P. H.; Investigation: all; Writing – Original Draft: F. L., O. P. H., Writing – Review and Editing: all; Visualization: F. L., Supervision, Project Administration, Funding acquisition: F. L., O. P. H., J. J.; Data Curation: B. Z., L. L., C. L., N. K., H. E., R. A., P. V., P. L.; Software: F. L., A. P., Y. Y., S. C., S. S., O. P. H.; **Competing interests:** Authors declare no competing interests; and **Data and materials availability:** All satellite data used in this work is publicly available through NASA and ESA web portals. GMI model output and policy response data is available upon request from the authors as is code to process all data sets.

**Materials and Methods**

Satellite $NO_2$ observations

We use retrieved $NO_2$ TVCD from both OMI and TROPOMI. OMI is a Dutch-Finnish UV-VIS spectrometer (*4*) on board the US National Aeronautics and Space Administration (NASA) Aura satellite that was launched in 2004. TROPOMI is a UV-VIS-NIR-SWIR instrument (*7*) on board the European Copernicus Sentinel-5 Precursor satellite that was launched in 2017. Both instruments similarly measure Earth radiance and solar irradiance spectra with spectral resolutions of approximately 0.5 nm. The ratio of radiance to irradiance at wavelengths between 400 and 496 nm is used to retrieve $NO_2$ TVCD. The ground footprint sizes are 13×24 $km^2$ and 3.5×5.5 $km^2$ (3.5×7 $km^2$ before August, 2019) at nadir for OMI and TROPOMI, respectively. Both instruments provide nearly daily to bi-daily global coverage with a local equator crossing times close to 13:30 h. We use the version 4.0 NASA OMI standard $NO_2$ products (*17*). We use the version 1.0.0 TROPOMI Level 2 offline $NO_2$ data products for 2019 and the version 1.1.0 data for 2020 (*18*). OMI and TROPOMI measurements are aggregated to resolutions of 0.25°×0.25° and 0.05°×0.05°, respectively. A given gridbox value is computed by averaging the pixel-level satellite observations weighted by the amount of the pixel footprint that overlaps the gridbox. We remove OMI observations with effective cloud fractions >30% to reduce retrieval errors and those affected by the so-called "row anomaly" (*19*). For TROPOMI, we use only observations with quality assurance values > 0.75.

For the maps shown, we calculate 20-day means of $NO_2$ TVCD around the Lunar New Year using OMI during 2015–2020 and TROPOMI for 2019 and 2020. We only include regions dominated by anthropogenic $NO_x$ emissions in the analysis; these are defined as regions with average annual OMI $NO_2$ TVCDs > 1×$10^{15}$ molec/$cm^2$ over the period of 2005–2019 (Fig S7;

*20*). For time series analysis, we further compute 7-day running averages to smooth out daily fluctuations in NO₂ TVCD due to retrieval noise, including the effects of clouds, and influences of meteorology (wind-driven transport influences NO₂ TVCDs).

Sector information

We select facilities with reported $NO_x$ emissions > 5 Gg/yr (*21*). The locations of 245 heavy industry plants including steel, iron, coke, oil, cement and glass industry, and 103 power plants considered in this analysis are shown in Figure S7. We compute 7-day running averages of OMI NO₂ TVCD for gridboxes where large power plants and other industrial plants are located for 2020 ($TVCD_{2020}$) and the mean of 2015–2019 ($\overline{TVCD_{2015-2019}}$). We calculate the relative difference as $(TVCD_{2020} - \overline{TVCD_{2015-2019}})/\overline{TVCD_{2015-2019}}$.

GEOS-CCM NO₂ simulations

We ran the GEOS-CCM (*9*) with anthropogenic and biomass burning emissions of $NO_x$ and other trace gas emissions held constant to simulate NO₂ TVCD over China in order to estimate the potential impact of meteorology on NO₂ TVCDs from January to February, 2020. The simulation uses the Global Modeling Initiative (GMI) chemistry mechanism (*22*) and the Goddard Chemistry Aerosol Radiation and Transport component of GEOS-5 (*23, 24*) to interact with the GMI chemistry. The simulation's meteorology is constrained by the Modern-Era Retrospective analysis for Research and Applications, Version 2 (MERRA2; *25*) assimilated meteorological data from the NASA Global Modeling and Assimilation Office (GMAO) GEOS-5 data assimilation system. The constant anthropogenic emissions are from the Representative Concentration Pathways (RCP) 6.0 scenario (*26*) for January 2019, downscaled to higher resolution using the Emissions Database for Global Atmospheric Research (EDGAR) version 4.3.2 (*27*) inventory. Constant biomass burning emissions are the January 2020 monthly mean from the Quick Fire Emissions Dataset version 2 (QFED2; *28*). This simulation includes 72 vertical levels at a spatial resolution of 0.25° (latitude and longitude) and a model time step of 7.5 minutes. We sample the model output only when and where there are valid satellite observations.

Statistical analysis of policy responses

For the policy evaluation, we make use of the timing of when the Chinese government first publicly reported that a person was infected with COVID-19, which occurred on several different dates across the country's provinces. The first public announcement of "viral pneumonia of unknown cause" in Wuhan occurred on January 3, 2020. Daily public health statements began on January 11, 2020, which included the new cases, deaths, and recoveries reported separately for each province. Of particular interest for our analysis are the times when the government announced the first case in *each* province (Table S1). We also use the exact timing when the government put restrictive mobility policies in place, in order to reduce the likelihood of transmission. The first such policy was put in place for Wuhan on January 23, 2020, followed by more restrictions for other provinces shortly after (Table S1).

We conduct a statistical evaluation of the exact timing of the reduction in $NO_2$ TVCDs. While the 2020 Lunar New Year coincided roughly with the lockdown of most Chinese provinces, the government's policy actions actually took two forms and varied over time. The first policy action was public announcements of new cases in each province, while the second policy action was to restrict movement and order citizens to stay in-doors (which became known as "lockdown"). We explore the timing of these two potential candidates—announcements of new cases and restrictive mobility policies—to identify to what extent they are responsible for NO$_2$ TVCD reductions. We take advantage of the temporal variation of these measures across the country.

To analyze the effects of these policies, we use fixed-effects models that predict tropospheric NO$_2$ TVCD, controlling for previous years' NO$_2$ TVCD as well as fixed effects for each province:

$$z_{t,p} = \alpha + \beta x_{t,p} + \delta z_{prior} + v_p + \varepsilon_{t,p} \tag{S1}$$

where $z$ is the outcome variable (daily NO$_2$ TVCD for the period from 4 weeks before LNY to 8 weeks after LNY), $x$ is an indicator variable on and after the first case is announced on day $t$ in province $p$ (which remains 1 after the first case; otherwise coded as 0), $z_{prior}$ is the NO$_2$ TVCD in prior years (which is the average of years 2015 and 2019 for the OMI data and of the year 2019 for the TROPOMI data where prior data is only available for 2019), $\alpha$ is the average fixed effect across all provinces and $v$ is the fixed effect of province $p$ (relative to $\alpha$), and $\varepsilon$ is an error term that is clustered at the province $p$.

To estimate the effect of the lockdown policy, we use the following fixed-effects model:

$$z_{t,p} = \alpha + \lambda y_{t,p} + \delta z_{prior} + v_p + \varepsilon_{t,p} \tag{S2}$$

where $y$ is an indicator variable for the lockdown of the province $p$ starting on day $t$ (which is 1 during the time of the lockdown; otherwise coded as 0), and all other variables are as defined above.

We use a similar fixed-effects model predicting the effect of both policies jointly:

$$z_{t,p} = \alpha + \beta x_{t,p} + \lambda y_{t,p} + \delta z_{prior} + v_p + \varepsilon_{t,p} \tag{S3}$$

where all variables are as previously specified. $\beta$, $\lambda$ and $\delta$ are the derived coefficients of the model.

Using the above specified fixed-effect models enables us to estimate the effect of the policy precisely, as we hold constant province-specific variation as well as prior year variation in $NO_2$. Our primary analysis uses OMI data (Table 1) but our results are qualitatively unchanged if we use TROPOMI data (Table S2).

Meteorology

Figure S6 shows the relative pre ($TVCD_{pre}$) to peri ($TVCD_{peri}$) period differences of $NO_2$ TVCD in 2020 for OMI (Fig S6A) and GEOS-CCM simulations with constant emissions (Fig S6B). We calculate the relative difference as $(TVCD_{peri} - TVCD_{pre})/TVCD_{pre}$. The relative changes from satellite observations for most areas in China are negative, with an average of -48%. The relative change of model-simulated (with constant emissions) TVCDs is 4% on average; simulations show a positive change over most regions over China.

We use the difference in relative changes (Fig S6C) to represent the changes in $NO_x$ emissions assuming that the relationship between $NO_2$ TVCDs on $NO_x$ emissions is linear. Considering that natural sources of $NO_x$ are significantly smaller than anthropogenic sources and do not change rapidly over such short time, the differences are then presumed to be primarily due to changes in anthropogenic $NO_x$ emissions. Decreases are widely observed, implying actual decreases in $NO_x$ emissions.

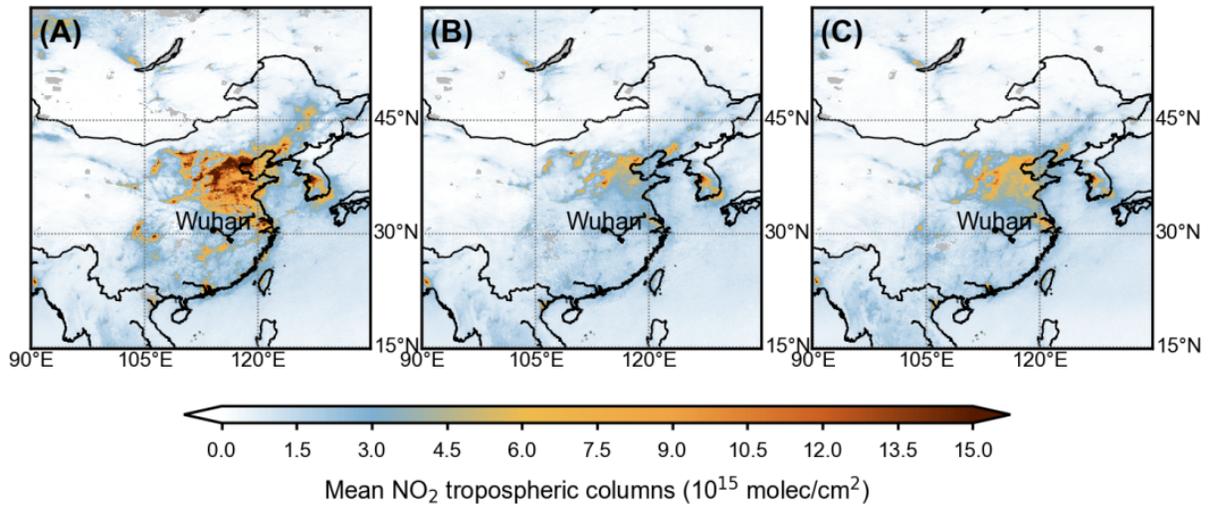

**Fig. S1.**

Similar Figure 1, but for TROPOMI. (A) -20 to -1, (B) 0-19, and (C) 20-39 days relative to the 2020 Lunar New Year.

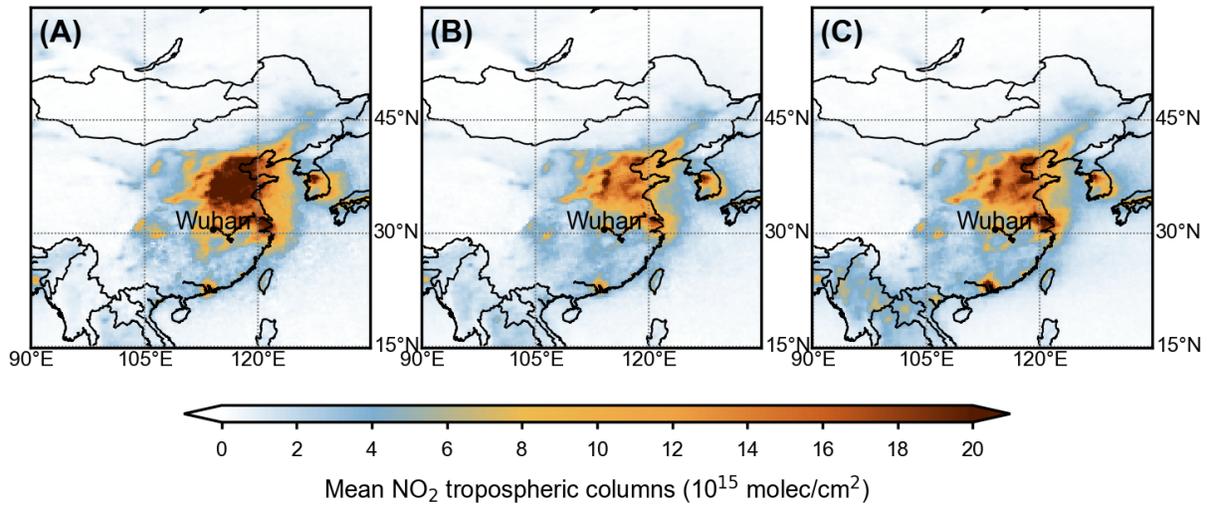

**Fig. S2.**

Similar to Figure 1, but for the OMI mean of 2005–2019.

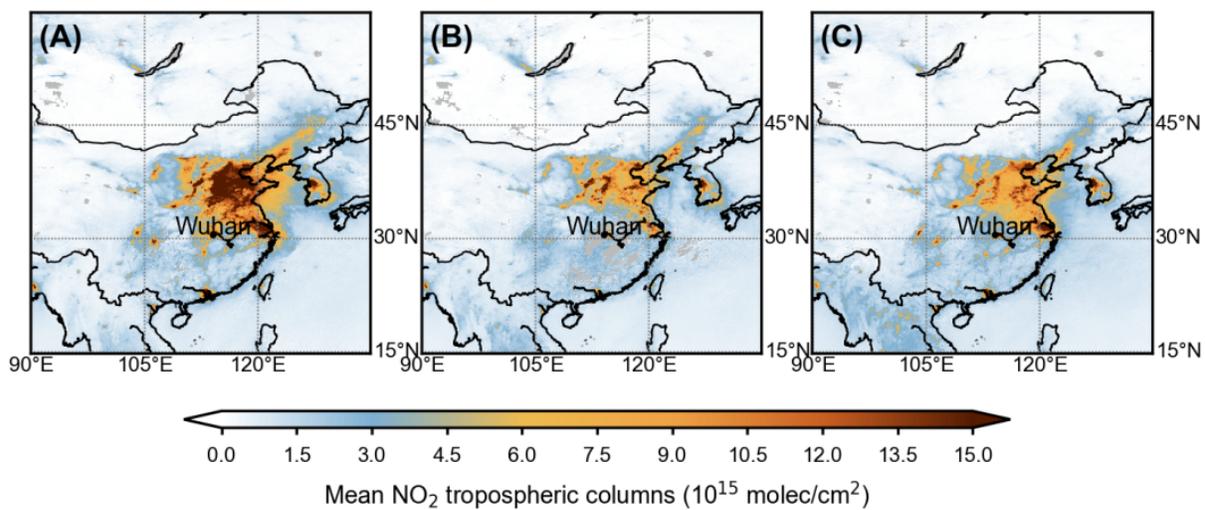

**Fig. S3.**

Similar to Figure 1, but for year 2019 (data from TROPOMI).

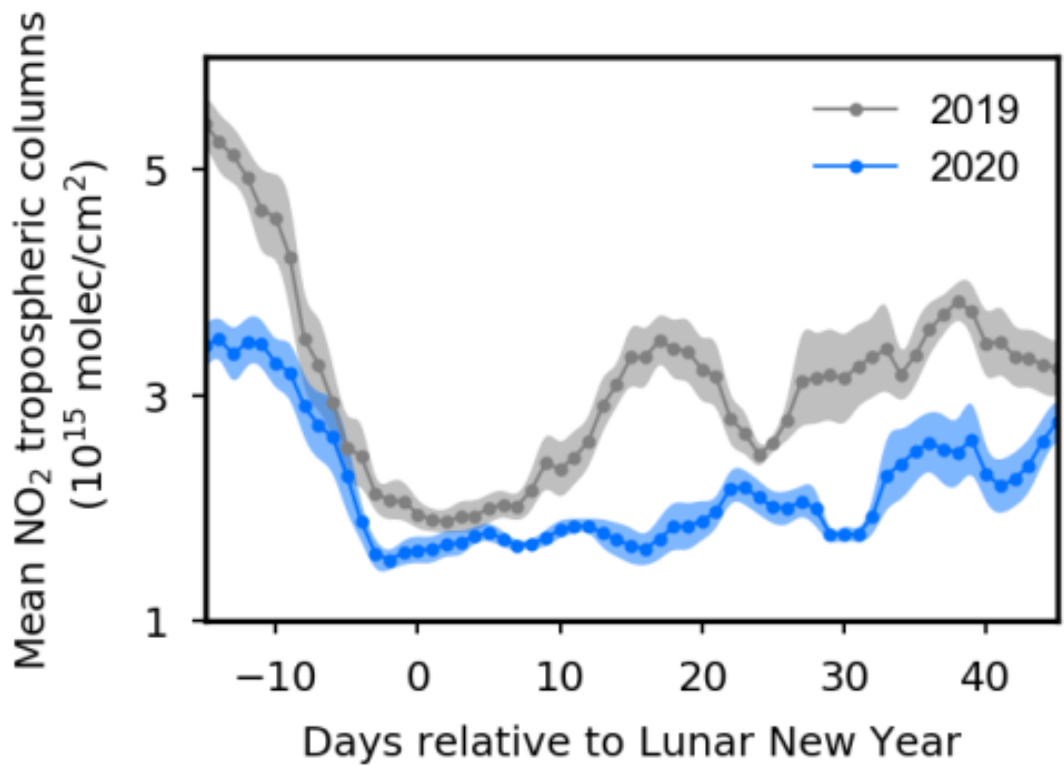

**Fig. S4.**

Similar to Figure 2, but for TROPOMI of 2019 and 2020.

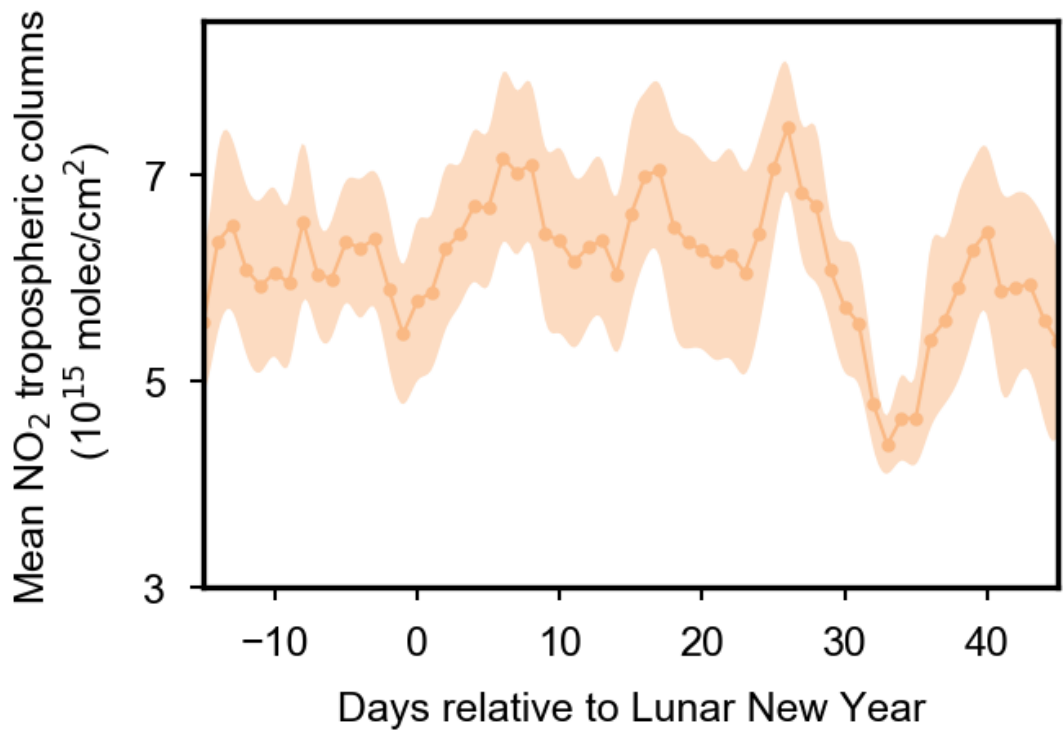

**Fig. S5.**

Similar to Figure 2, but for GEOS-CCM simulation with constant emissions for 2020.

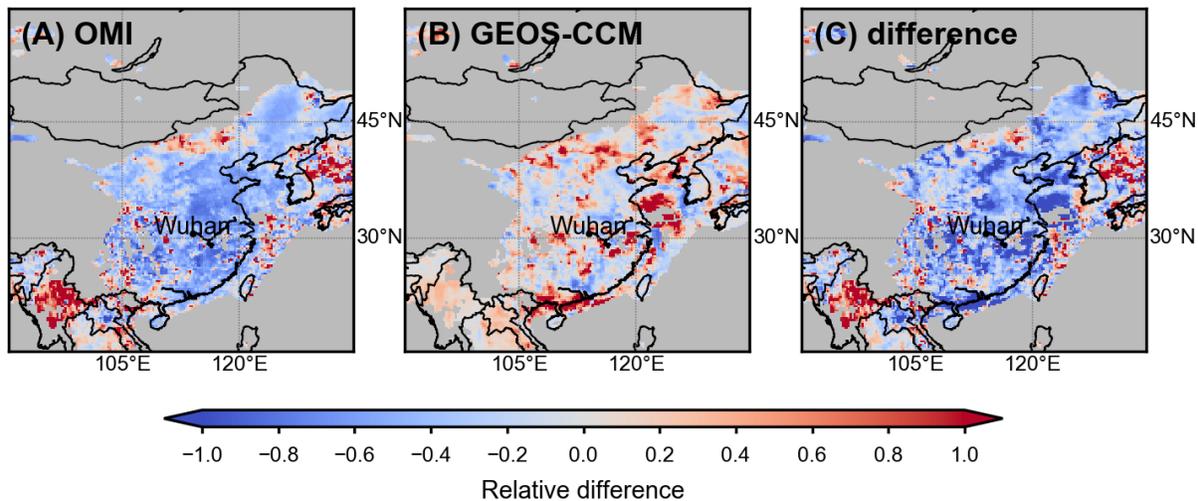

**Fig. S6.**

Relative difference of NO$_2$ TVCD from the pre to peri period from (A) OMI, (B) GEOS-CCM simulation with constant emissions, (C) and their difference. We define regions where NO$_x$ emissions are not dominated by anthropogenic sources as those with average annual OMI NO$_2$ TVCDs < 1×10$^{15}$ molec/cm$^2$ over the period of 2005–2019, shown in gray.

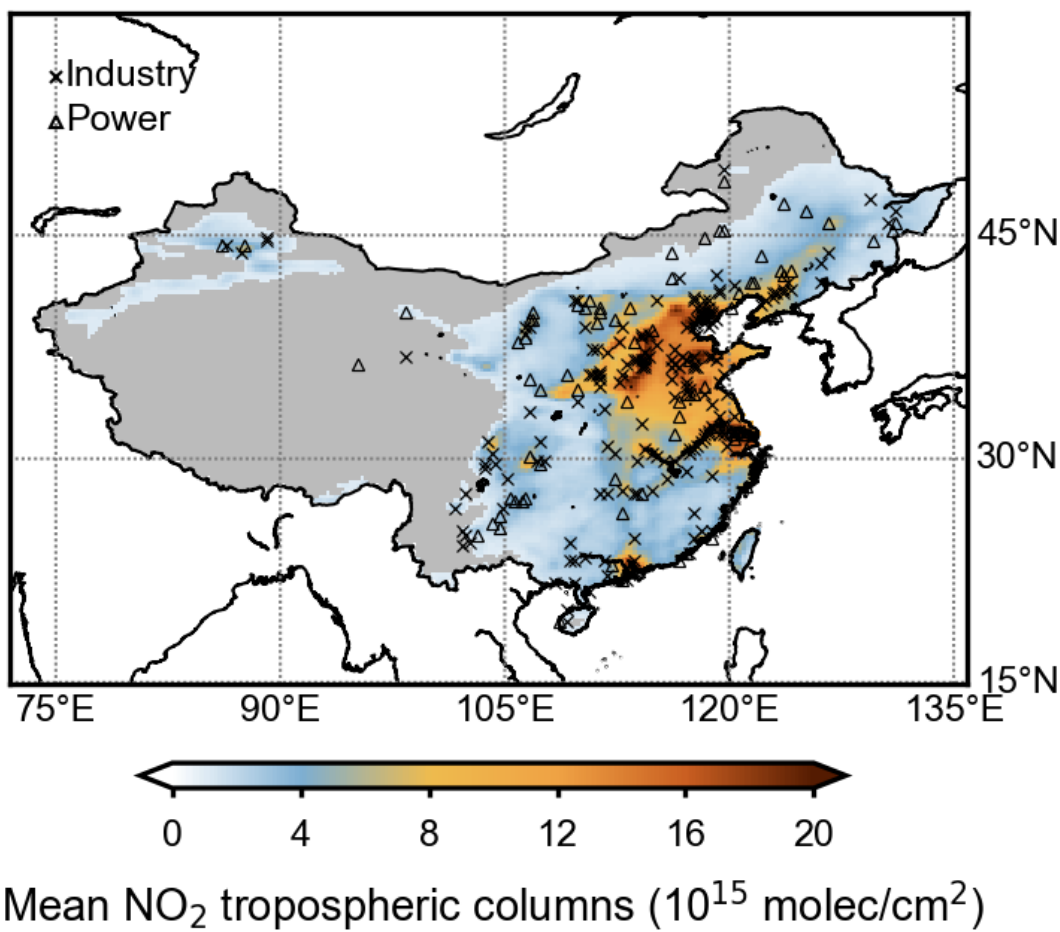

**Fig. S7.**

The locations of selected industry and power plants. The background is the OMI NO$_2$ TVCD over China average for 2005–2019. We define regions where NO$_x$ emissions are not dominated by anthropogenic sources as those with average annual OMI NO$_2$ TVCDs < 1×10$^{15}$ molec/cm$^2$ over the period of 2005–2019, shown in gray.

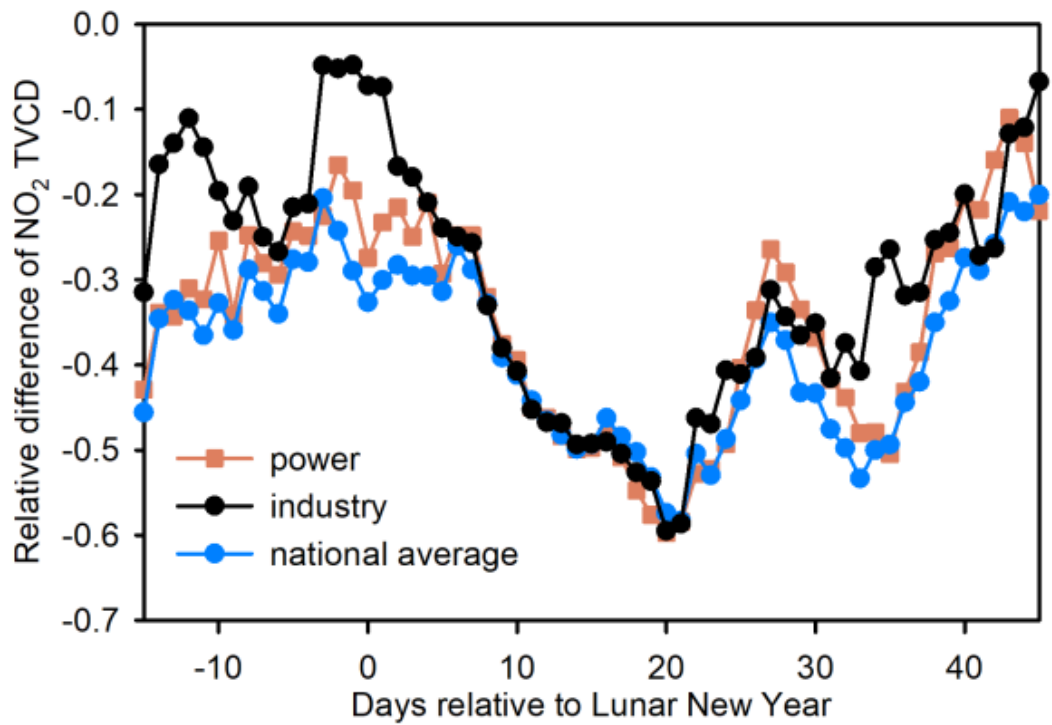

**Fig. S8.**

Relative difference of 7-day moving averages of OMI NO$_2$ TVCDs between 2020 and the mean of 2015–2019.

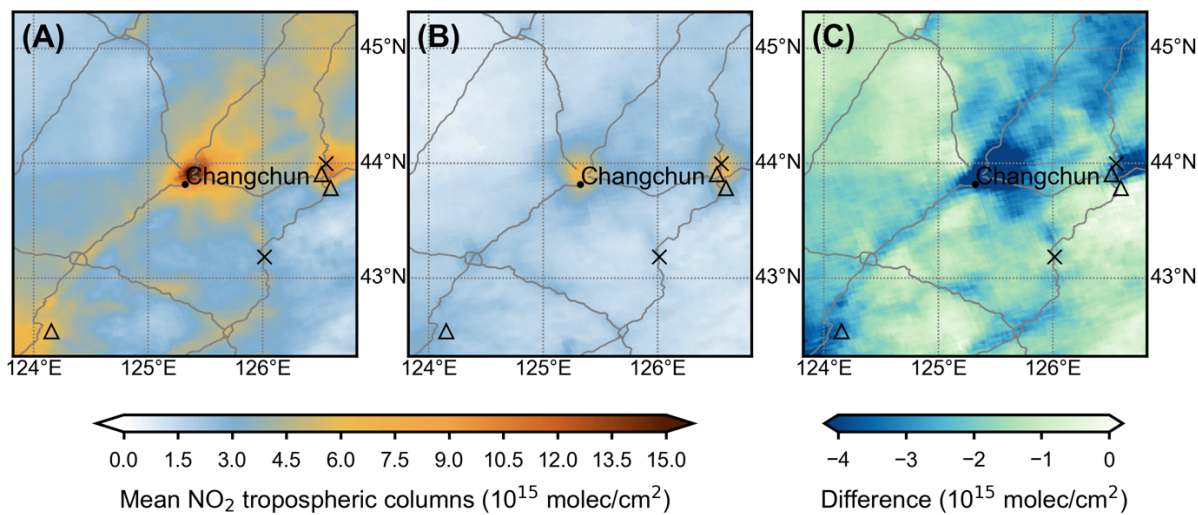

**Fig. S9.**

Average TROPOMI NO$_2$ TVCD over Changchun, China (black dot) for 20 days (A) prior to and (B) after the 2020 Lunar New Year, and (C) their difference. The locations of large power plants and other industrial plants are indicated by triangle and x, respectively. The lines show China National Highways.

**Table S1.**

A timeline of the date of the first case reported and the date the government put restrictive policies in effect, by province.

| Province | Date of first case reported | Date of government response |
|---|---|---|
| Hubei | 03/01/2020 | 23/01/2020 |
| Beijing | 18/01/2020 | 24/01/2020 |
| Sichuan | 18/01/2020 | 25/01/2020 |
| Guangdong | 19/01/2020 | 23/01/2020 |
| Shanghai | 20/01/2020 | 25/01/2020 |
| Chongqing | 21/01/2020 | 25/01/2020 |
| Henan | 21/01/2020 | 25/01/2020 |
| Hunan | 21/01/2020 | 25/01/2020 |
| Jiangxi | 21/01/2020 | 25/01/2020 |
| Shandong | 21/01/2020 | 25/01/2020 |
| Yunnan | 21/01/2020 | 25/01/2020 |
| Zhejiang | 21/01/2020 | 23/01/2020 |
| Anhui | 22/01/2020 | 25/01/2020 |
| Fujian | 22/01/2020 | 25/01/2020 |
| Guangxi | 22/01/2020 | 25/01/2020 |
| Guizhou | 22/01/2020 | 25/01/2020 |
| Hebei | 22/01/2020 | 25/01/2020 |
| Jiangsu | 22/01/2020 | 25/01/2020 |
| Liaoning | 22/01/2020 | 25/01/2020 |
| Ningxia | 22/01/2020 | 25/01/2020 |
| Shanxi | 22/01/2020 | 25/01/2020 |
| Gansu | 23/01/2020 | 25/01/2020 |
| Heilongjiang | 23/01/2020 | 25/01/2020 |
| Jilin | 23/01/2020 | 25/01/2020 |
| Shaanxi | 23/01/2020 | 25/01/2020 |
| Xinjiang | 23/01/2020 | 25/01/2020 |
| Inner Mongolia | 24/01/2020 | 25/01/2020 |
| Qinghai | 25/01/2020 | 25/01/2020 |
| Tianjin | 25/01/2020 | 26/01/2020 |

**Table S2.**

Effects of the government policies on $NO_2$ TVCD.

|  | Outcome variable: | | |
|---|---|---|---|
|  | $NO_2$ TVCD ($10^{15}$ molec/cm$^2$) | | |
|  | (1) | (2) | (3) |
| First case announced in province, $\beta$ | -1.869*** |  | -1.462*** |
|  | (0.368) |  | (0.334) |
| Lockdown of province, $\lambda$ |  | -1.242*** | -0.568*** |
|  |  | (0.228) | (0.111) |
| $NO_2$ TVCD 2019, $\delta$ | 0.025** | 0.032*** | 0.023** |
|  | (0.008) | (0.008) | (0.008) |
| Constant, $\alpha$ | 4.136 | 3.307 | 4.167 |
| Number of observations | 1,275 | 1,275 | 1,275 |
| $R^2$ | 0.567 | 0.543 | 0.575 |
| Adjusted $R^2$ | 0.557 | 0.532 | 0.565 |

*Note*. $NO_2$ TVCD is based on TROPOMI. We use a fixed-effects model (Eqs. S1-3) with first case announced and lockdown coded as binary indicator variables. We control for the previous year's $NO_2$ TVCD to adjust for seasonal variation and include provinces' fixed-effects to adjust for geographical variation. The "Constant" term is the average province fixed-effect used as a baseline to compare the relative effect of the policy interventions. All standard errors (shown in parentheses) are clustered at the province level. * $p < 0.05$, ** $p < 0.01$, *** $p < 0.001$.